\documentclass[a4paper,11pt]{article}
\pdfoutput=1
\usepackage{datetime}
\usepackage[utf8]{inputenc}
\usepackage{jheppub} 
\usepackage{color}
\usepackage{graphicx, multirow,soul,amsmath,amsfonts,amssymb,mathrsfs,amsfonts}
\usepackage[all]{hypcap}
\usepackage{url}
\usepackage{bbm}
\usepackage{cancel}
\usepackage{slashed}
\usepackage[dvipsnames]{xcolor}
\usepackage{multicol, blindtext}
\usepackage{lipsum}
\usepackage{mattens}
\usepackage{mathtools}
\usepackage{amsbsy}
\usepackage{epsfig}
\usepackage{footnote}
\usepackage{float}
\usepackage{cleveref}
\usepackage{caption}
\usepackage{subcaption}
\usepackage{hyperref}
\hypersetup{
    linktocpage=true,
    colorlinks=true,
    allcolors=blue
}

\numberwithin{equation}{section}

\newcommand{\iee}{{\it i.e.}}

\newcommand{\mbino}{\ensuremath{M_{\widetilde{B}}}}
\newcommand{\octino}{\mathcal{O}}
\newcommand{\mGr}{\ensuremath{m_{3/2}}}
\newcommand{\tB}{\widetilde{B}}
\newcommand{\tW}{\widetilde{W}}

\newcommand{\tg}{\widetilde{g}}

\newcommand{\uc}{\mathbf{u}}
\newcommand{\vc}{\mathbf{v}}
\newcommand{\bivo}{bi$\nu$o}
\newcommand{\wivo}{wi$\nu$o}
\title{A Hybrid Type I + III Inverse Seesaw Mechanism in $U(1)_{R-L}$-symmetric MSSM}

\author[a,b]{Cem Murat Ayber,}
\author[a]{Seyda Ipek}

\affiliation[a]{Department of Physics, Carleton University,

 1125 Colonel By Drive, Ottawa, ON, Canada}
\affiliation[b]{Arthur B. McDonald Canadian Astroparticle Physics Research Institute, Queen's University, 

64 Bader Lane, Kingston, ON, Canada}
\emailAdd{cemayber@cmail.carleton.ca}
\emailAdd{sipek@physics.carleton.ca}

\abstract{
We show that, in a $U(1)_{R-L}$-symmetric supersymmetric model, the pseudo-Dirac bino and wino can give rise to three light neutrino masses through effective operators, generated at the messenger scale between a SUSY breaking hidden sector and the visible sector. The neutrino--bino/wino mixing follows a hybrid type I+III inverse seesaw pattern. The light neutrino masses are governed by the ratio of the $U(1)_{R-L}$--breaking gravitino mass, $m_{3/2}$, and the messenger scale $\Lambda_M$. The charged component of the $SU(2)_L$--triplet, here the lightest charginos, mix with the charged leptons and generate flavor-changing neutral currents at tree level. We find that resulting lepton flavor violating observables yield a lower bound on the messenger scale, $\Lambda_M \gtrsim (500-1000)~{\rm TeV}$ for a simplified hybrid mixing scenario. We identify interesting mixing structures for certain $U(1)_{R-L}$--breaking singlino/tripletino Majorana masses. For example, in some parameter regimes, bino or wino has no mixing with the electron neutrino. We also describe the rich collider phenomenology expected in this neutrino-mass generation mechanism. 

}

\begin{document}
\maketitle
\flushbottom

%%%%%%%%%%%%%%%%%%%%%%%%%%%%%%%%%%%
%%%%%%%% Introduction %%%%%%%%%%%%%
%%%%%%%%%%%%%%%%%%%%%%%%%%%%%%%%%%%
\section{Introduction}\label{Section:intro}
%%%%%%%%%%%%%%%%%%%%%%%%%%%%%%%%%%%

One of the triumphs of contemporary physics is the observation of neutrino oscillations, which provides direct evidence that at least two of the neutrinos in the Standard Model (SM) have mass. Current efforts of the neutrino physics experiments yield the neutrino oscillation observables to be \citep{Esteban:2020nufit,SajjadAthar:2021prg,Caravaca:2020xip}
\begin{equation}
    \begin{aligned}
      &  \Delta m_{12}^2 = 7.41^{+0.21}_{-0.20}\times 10^{-5}\ {\rm eV^2},\quad |\Delta m_{13}^2| = 2.507^{+0.026}_{-0.027}\times 10^{-3}\ {\rm eV^2}, \\
&\sin^2 \theta_{12}= 0.303^{+0.012}_{-0.012}\, ,\hspace{0.5cm} \sin^2 \theta_{23}= 0.451^{+0.019}_{-0.016}\, ,\hspace{0.5cm}  \sin^2 \theta_{13}= 0.02225^{+0.00056}_{-0.00059}.
    \end{aligned}
    \label{eqn:numixvariables}
\end{equation}
It is well-known that explaining the origin of neutrino masses requires going beyond the SM. Thus, it is reasonable to seek a natural way to explain such a phenomenon. Over the past several decades, a plethora of mechanisms have been introduced to explain neutrino masses. By far, the most popular approach is the seesaw mechanism where the light neutrino masses are inversely proportional to a heavy scale associated with a right-handed (RH) neutrino mass (e.g. ~see \citep{KING:2016456,Cai:2018,Xing:2021} and the references therein). Specifically, in the type I seesaw mechanism, the RH neutrinos are SM singlets, while in the type III seesaw mechanism, they are $SU(2)_L$-triplets. On the other hand, in the inverse-seesaw (ISS) mechanism, the RH neutrinos are pseudo-Dirac particles and the light neutrino masses are \emph{proportional} to a small Majorana mass~ \citep{WYLER:1983205,Mohapatra:1986aw,Mohapatra:1986bd}. In ISS, lepton number is approximately conserved, broken only by the small Majorana masses of the RH neutrinos, providing a natural way to explain the lightness of the SM neutrinos. 

In \citep{Coloma:2016vod} it was shown that pseudo-Dirac binos in a $U(1)_R$-symmetric minimal supersymmetric SM (MSSM) can act like RH neutrinos and generate the light neutrino masses.  The pseudo-Dirac bino, called \bivo~, is an SM singlet;hence, this mechanism is equivalent to a type I ISS scenario. In this scenario, the small Majorana bino mass is proportional to the gravitino mass $\mGr$, which breaks the $U(1)_R$ global symmetry. In turn, the light neutrino masses are proportional to $\mGr v^2/\Lambda_M^2$, where $v$ is the Higgs vacuum expectation value (vev) and $\Lambda_M$ is the messenger scale between a SUSY--breaking hidden sector and the visible sector. (The model is described in more detail in the next section.) Importantly, in \citep{Coloma:2016vod} the lightest neutrino is predicted to be massless since \bivo ~has only two degrees of freedom.

In $U(1)_R$-symmetric MSSM, all gauginos are pseudo-Dirac fermions. In this paper, we investigate the scenario where the wino, as an $SU(2)_L$-triplet, is also involved in neutrino-mass generation. In this way, $U(1)_R$-symmetric MSSM naturally gives rise to a hybrid type I+III ISS mechanism.\footnote{Such hybrid seesaw mechanisms also arise in, e.g., the adjoint $SU(5)$ models~\citep{FileviezPerez:2007bcw,FileviezPerez:2007yji} and in gauged $B-L$ MSSM. We emphasize that these models require one massless neutrino whereas the model we describe here can accommodate three massive neutrinos in its most general form. Furthermore, while these models generate the neutrino masses through a seesaw mechanism, in our case both the type I and type III are \emph{inverse} seesaws. This ISS texture affects the low energy constraints, as discussed later in the text.} In addition to strongly motivating a hybrid texture for the neutrino-mass generation, this extension makes it possible to give mass to all three neutrinos and provides a much richer phenomenology at the LHC, which we discuss here.

The organization of this paper is as follows. 
In \Cref{Section:model}, we summarize the relevant parts of the $U(1)_{R-L}$--symmetric MSSM. In \Cref{sec:neutrinomasses}, we explain the neutrino-mass generation mechanism in detail. In \Cref{Section:LFVconst}, we derive the low energy constraints on this scenario coming from lepton-flavor violating observables. In \Cref{Section:LHCpheno}, we discuss the interesting phenomenology expected to be seen at the LHC. We conclude in \Cref{sec:conclusions}.

%%%%%%%%%%%%%%%%%%%%%%%%%%%%%%%%%%%
%%%%%%%% Model %%%%%%%%%%%%%%%%%%%%
%%%%%%%%%%%%%%%%%%%%%%%%%%%%%%%%%%%
\section{Model}\label{Section:model}
%%%%%%%%%%%%%%%%%%%%%%%%%%%%%%%%%%%

In this work, we study a $U(1)_R$-symmetric SUSY model. In these models, a global $U(1)_R$ symmetry is imposed on the supersymmetric sector such that the superpartners have $+1$ $R$-charges while the SM fields are neutral under $U(1)_R$. In contrast to MSSM, due to this charge assignment, gauginos in $U(1)_R$-symmetric SUSY cannot be Majorana particles. Dirac gaugino masses, on the other hand, can be generated by introducing three adjoint superfields with $-1$ $R$-charges: a hypercharge singlet  $S$, an $SU(2)_L$ triplet  $T$, and an $SU(3)_c$ octet $\octino$ \citep{Fox:2002bu}. Furthermore, $U(1)_R$ symmetry also forbids the Higgsino mass terms. In order to give mass to Higgsinos, two inert doublets $R_{u,d}$ are introduced.\footnote{As shown in \citep{Fok:2012fb,Bertuzzo:2015dg}, 125 GeV Higgs can be accommodated in such scenarios.} (Relevant superfields and their charge assignments are given in \Cref{tab:fields}.) The terms that are forbidden by the $U(1)_R$ symmetry in these models help to alleviate the SUSY flavor and $CP$ problems~\citep{Kribs:2007ac}. Furthermore, heavier stops require less Higgs fine-tuning in SUSY models with $U(1)_R$ symmetry. This is especially important given the stringent LHC limits on stops and gluinos, which negate the MSSM solution to the hierarchy problem \citep{Aad:2021gluino,Sirunyan:2021stop}.

 In order to generate the neutrino masses in a minimal way in these SUSY models, we extend the $U(1)_R$ symmetry to $U(1)_{R-L}$, where $L$ is the lepton number. (See \Cref{tab:fields}.) In this case, the SM leptons are charged under the global $U(1)_{R-L}$, which allows the mixing between the neutralinos and the active neutrinos. (see also \citep{Bertuzzo:2012su,Frugiuele:2013rpv} for a similar approach.) Therefore, the $U(1)_{R-L}$ symmetry can provide a natural mechanism for neutrino mass generation. 
\begin{table}
\centering
\begin{tabular}{|c|c|c|c|c|}
\hline
Superfields & $SU(2)_L$ & $U(1)_Y$	&	$U(1)_R$	&	$U(1)_{R-L}$ \\
\hline\hline
%$Q, U^c, D^c$	&	 1	&	1 \\
$L_{i}$	& 2& -1/2	&1	&	0 \\
$E_{i}^c$  &1 &1	&	1	&	2	\\	
\hline
$H_{u,d}$  & 2 & 1/2	&	0	&	0	\\
$R_{u,d}$	& 2 & -1/2 &	2	&	2	\\
\hline
$W_{\widetilde{B}}^\alpha $ & 1 & 0	&	1	&	1	\\
$\Phi_S$ &1 &0	&	0	&	0	\\
$W_{\widetilde{W}}^\alpha$ & 3 & 0	&	1	&	1	\\
$\Phi_T$ &3 &0	&	0	&	0	\\
\hline
$\phi $ & 1 &0 & 1 &  1 \\ 
\hline
$W'_\alpha $  & 1 &0 & 1 &  1 \\
\hline
\end{tabular}
\caption{The relevant superfield content of the model and their charge assignments. The $R-L$ charge is defined as the $U(1)_{R}$-charge minus the lepton number of the field. $L_i,\ E_{i}^{c}$ are the lepton superfields and the subindex \textit{i} indicates the fermion generation. The fermionic components of the superfields $R_{u,d}$ are the Dirac partners of the Higgsinos $H_{u,d}$. $\Phi_{S, T}$ are superfields that have the same SM charges as $W^{\alpha}_{\tB,\tW}$ and their fermionic components, $S,~T$ are the Dirac partners of the bino and the wino, respectively. $\phi$ is the conformal compensator and $W^\prime_{\alpha}$ is a $D$--term spurion field.}\label{tab:fields}
\end{table}

We incorporate SUSY breaking via $F$-- and $D$--terms. We consider two independent sectors that break SUSY. The first sector, $\sqrt{F_1},D_1\sim 10~$TeV, is coupled to the superfields whereas the second sector, $\sqrt{F_2},D_2\gtrsim 10^4~$TeV, is coupled to the SM only through gravity. (These scales are motivated by the resulting phenomenology~\citep{Gehrlein:2019gtk}.) SUSY breaking is communicated to the visible sector at a messenger scale $\Lambda_M$. As shown in \citep{Fox:2002bu}, supersoft SUSY breaking, \iee~emergence of Dirac gauginos, is incorporated by the  spurion field $W'_{\alpha}=\theta_\alpha D$, where $D=\langle W'_\alpha \rangle$  is the  SUSY--breaking vev of a \textit{D}--term spurion field. Therefore, Dirac masses of the gauginos are generated through the following supersoft operator by integrating out the messenger field at a scale $\Lambda_M$~\citep{Fox:2002bu}, 
\begin{equation}
\label{eqn:supersoftop}
\int d^2\theta\, \frac{\sqrt{2} c_j}{\Lambda_M} W'_\alpha W_{\widetilde{\chi}_j}^\alpha\Phi_{\eta_j} \Longrightarrow \frac{\sqrt{2} c_j}{\Lambda_M} D (\widetilde{\chi}\,\eta)_j \equiv M_{\tilde{\chi}_j}(\widetilde{\chi}\,\eta)_j ~, \quad \left ( \widetilde{\chi}\, \eta\right )_{j} =  \tB S \,,\,   \tW T \,,\,  \tg \octino\,,
\end{equation}
where $c_j$ is a dimensionless coefficient that we take to be $\mathcal{O}(1)$, $\Phi_{\eta_j}$ is the chiral superfield whose fermionic component is the Dirac partner of the relevant gaugino $ \widetilde{\chi}_j$, and index-\textit{j} ensures the correct gaugino-Dirac partner pair. In particular, $\widetilde{\chi}_j$ and $\eta_j$ are the Weyl components of the Dirac field $\psi_j^{T}=(\widetilde{\chi}_j,\eta_j^{\dagger})^T$. 
We assume that scalar adjoints only receive a finite soft mass from the operators in \Cref{eqn:supersoftop} at one loop (for details, see, e.g., ~\citep{Fox:2002bu}). We note that although supersoft terms of the form $W'_{\alpha}W'^{\alpha}\Phi_{\eta_j}\Phi_{\eta_j}$ will generally contribute to the masses of the scalar adjoints, these soft terms can be forbidden \citep{Alves:2015gg}.

As with all global symmetries, $U(1)_{R-L}$ is broken due to gravity. Consequently, Majorana masses of the gauginos are generated via anomaly mediation \citep{Randall:1998uk,Giudice:1998xp,Gherghetta:1999am},
\begin{align}
m_{\widetilde{\chi}_j}=\frac{\beta(g_{\widetilde{\chi}_j})}{g_{\widetilde{\chi}_j}}m_{3/2},
\end{align}
where $\beta(g_{\widetilde{\chi}_j})$ are the beta functions of the relevant SM gauge coupling $g_{\widetilde{\chi}_j}$, and $\mGr$ is the gravitino mass. The gravitino picks up mass from all sources of SUSY breaking, $m_{3/2} = \sqrt{ \widetilde{F}_{1}^2 + \widetilde{F}_{2}^2}/ \sqrt{3} M_{\rm Pl}$ where $M_{\rm Pl}$ is the Planck mass and $\tilde{F}_i^2=F_i^2 + D_i^2/2$. We emphasize that $U(1)_{R-L}$ is approximately conserved as long as the gravitino $m_{3/2}$ is light. Hence, we assume the messenger scale $\Lambda_M$ is below the Planck scale, $\Lambda_{M}\ll M_{\rm Pl}$ which leads to a large hierarchy between Dirac and Majorana gaugino masses, $m_{\widetilde{\chi}_j}\ll M_{\widetilde{\chi}_j}$. Furthermore, $U(1)_{R}$-breaking Majorana masses for the Dirac partners could also be generated. We assume that the Majorana masses of the Dirac partners are also proportional to the gravitino mass and can be parameterized as 
\begin{equation}
  m_{S}\equiv \kappa_S\, m_{3/2},\quad \, m_{T}\equiv \kappa_T\, m_{3/2},
  \label{eqn:STMajoranamass}
\end{equation} 
where $\kappa_S$ and $\kappa_T$ are dimensionless coefficients. 
Thus, the pair $\psi_j^{T}=(\widetilde{\chi}_j,\eta_j^{\dagger})^T$ becomes a pseudo-Dirac fermion. We will drop the index-\textit{j} and refer to pseudo-Dirac bino--singlino and wino--tripletino pairs as bi$\nu$o and wi$\nu$o as they will play a role in neutrino mass generation and mixing. Gluino and octino will be ignored since they are not relevant to this work. 

After electroweak symmetry breaking (EWSB), adjoint fermions \textit{S} and \textit{T} participate in both neutralino and chargino mixing due to the presence of $U(1)_R$ symmetry. The relevant part of the superpotential that accounts for both neutralino and chargino mixing is given as~\citep{Kribs:2008hq}
\begin{align}\label{eqn:Wneutralino}
\mathcal{W}=\mu_u H_u R_u + \mu_d H_d R_d + \Phi_S \left( \lambda^u_{\widetilde{B}} H_u R_u + \lambda^d_{\widetilde{B}} H_d R_d  \right) + \Phi_T \left( \lambda^u_{\widetilde{W}} H_u R_u + \lambda^d_{\widetilde{W}} H_d R_d  \right).
\end{align}
We work in the large $\tan\beta \equiv {v_u}/{v_d}$ limit, $v_d \to 0$, where $v_{u,d}\equiv \langle H_{u,d}^0\rangle$ are the up/down-type Higgs vevs with $v_u^2+v_d^2=v^2/2 \simeq(174~{\rm GeV})^2$.
In this limit, \Cref{eqn:Wneutralino} generates the neutralino mixing matrix 
\begin{align}
\mathbb{M}_N\simeq\begin{pmatrix}
		M_{\widetilde{B}}	&	0	&	{g_Y v}/{2}	&	0 \\
		0			&	M_{\widetilde{W}}	&	-{g_2 v}/{\sqrt{2}}		&	0 \\
		{\lambda_{\widetilde{B}}^u v}/{2}	&	-{\lambda_{\widetilde{W}}^u v}/{2}	&	\mu_u	&	0	\\
		0	&	0	&	0	&	\mu_d
\end{pmatrix},
\end{align}
in the basis $( \widetilde{B},\widetilde{W}^0, \widetilde{R}_u^0, \widetilde{R}_d^0 )\times ( S,T^0,\widetilde{h}_u^0, \widetilde{h}_d^0 )$. One state with mass $\mu_d$ is already decoupled for $v_d\to 0$. We further assume $\lambda^u_{\tB,\tW}=0$ such that \bivo,~\wivo ~and Higgsinos do not mix. Since possible mixing of neutralinos does not bring in any new degrees of freedom to the system, this assumption does not affect the neutrino mass generation scenario described in the next section. Neutralino mixing, however, is important for LHC phenomenology as discussed in \Cref{Section:LHCpheno}. Similarly, for only LHC phenomenology purposes, we will assume a hierarchy $\mu_{u}\approx \mu_{d} >M_{\widetilde{B},\widetilde{W}}$ and take the Higgsinos to be decoupled in the mass spectrum. This assumption is less motivated theoretically, as naturalness favors light Higgsinos.

Under the assumptions given above, the chargino mixing matrix is also almost diagonal,
\begin{align}
\mathbb{M}_{\rm C}\simeq\begin{pmatrix}
		M_{\tW}	&	-g_2 v/\sqrt{2}&0 	 \\
		0 & \mu_u & 0 \\
		0&	0	&	\mu_d
\end{pmatrix}\,,
\label{eqn:charginoM}
\end{align}
in the basis $( \tW^+,\widetilde{R}_u^+, \widetilde{R}_d^+) \times ( \Phi_T^-,\widetilde{h}_u^-, \widetilde{h}_d^- )$. We assume that the lightest charginos $\chi_1^\pm$ are purely composed of charged weakino states and that they are degenerate with the \wivo. In the next section, we will describe a type III ISS mechanism where the \wivo ~acts like an RH triplet. This will introduce mixing between weak charginos and charged leptons, which we will discuss in \Cref{Section:LFVconst}. 

%%%%%%%%%%%%%%%%%%%%%%%%%%%%%%%%%%%
%%%%%%%% Neutrino Masses %%%%%%%%%%
%%%%%%%%%%%%%%%%%%%%%%%%%%%%%%%%%%%
\section{Neutrino masses} \label{sec:neutrinomasses}
The $U(1)_{R-L}$-symmetric MSSM has a pseudo-Dirac SM-singlet fermion, namely the \bivo, and a pseudo-Dirac $SU(2)_L$-triplet fermion, the \wivo. In \citep{Coloma:2016vod}, it was shown that the pseudo-Dirac \bivo ~could generate neutrino masses through an ISS mechanism. In this minimal scenario, the lightest neutrino is predicted to be massless. We extend this framework to include the \wivo ~in generating the neutrino masses. These new degrees of freedom allow for all three neutrinos to be massive, open up more parameter space of interest, and provide a richer phenomenology at the LHC.

Building on \citep{Coloma:2016vod}, we introduce the following $U(1)_{R-L}$-conserving, dimension-6 operators:
\begin{equation}
\begin{aligned}
&\frac{1}{\Lambda_M^2}\int d^2\theta\,\left( f^{\, i}_{\tB} W'_\alpha W_{\tB}^\alpha H_u L_i + f^{\, i}_{\tW} W'_\alpha W_{\tW}^\alpha H_u L_i \right) \Longrightarrow  f^{\, i}_{\tB}\,  \frac{M_{\tB}}{\Lambda_M}\tB \, h_u \ell_i + f^{\, i}_{\tW}\,  \frac{M_{\tW}}{\Lambda_M}\tW \, h_u \ell_i\,,   
\label{eqn:numassgenBW}
\end{aligned}
\end{equation}
where $f^{\, i}_{\tB,\tW}$ are dimensionless coefficients and $i=e,\mu,\tau$.\footnote{Note that if bino, wino and higgsinos mix, the coefficients $f^i_{\tilde{B},\tilde{W}}$ are rescaled by a mixing angle. This will not affect the neutrino mixing structure.}  
In addition the $U(1)_{R-L}$-conserving terms above, we also introduce the following $U(1)_{R-L}$-breaking, dimension-5 terms:
\begin{equation}
\begin{aligned}
&\frac{1}{\Lambda_M}\int d^{2}\theta d^{2}\bar{\theta}\phi^\dagger\left( d^{\, i}_S\Phi_S H_u L_i + d^{\, i}_T\Phi_T H_u L_i\right) \Longrightarrow  \frac{m_{3/2}}{\Lambda_M} \left(d^{\, i}_S S \, h_u \ell_i  + d^{\, i}_T T \, h_u \ell_i\right)\, ,
\label{eqn:RLviol}
\end{aligned}
\end{equation}
where $d^{\, i}_{S/T}$ are dimensionless coefficients and $\phi=1+\theta^2 m_{3/2}$ is the conformal compensator. While the operators in \Cref{eqn:numassgenBW} explicitly violate R-parity and lepton number, they conserve $U(1)_{R-L}$. Moreover, the $U(1)_{R-L}$-violating terms in \Cref{eqn:RLviol} are highly suppressed compared to the $U(1)_{R-L}$-conserving terms because $\mGr \ll M_{\tB/\tW}$. We note in passing that although the effective operators in \Cref{eqn:RLviol} are introduced for completeness, unlike the original scenario, they are not required to generate the correct neutrino spectrum. We will discuss the details shortly.

After EWSB, the interactions in \Cref{eqn:numassgenBW,eqn:RLviol} generate mixing between neutrinos, \bivo ~and \wivo. We emphasize that the mixing of active neutrinos and bi$\nu$o has a type I ISS texture, whereas the mixing follows a type III ISS pattern for the wi$\nu$o case.\footnote{The parallelism between the mixing structure of seesaw scenarios mediated by fermionic singlets and triplets is discussed in, e.g.,~\citep{Abada:2007ux}.} Hence, this is a natural scenario that gives rise to a hybrid type I+III ISS mechanism in explaining the neutrino masses.

Using the prescription given in \citep{Gavela:2009cd}, neutrino mass matrix in the $(\nu_{i},\tB,\tW,S,T)$ basis after EWSB is
\begin{equation}
\mathcal{M}_\nu= \begin{pmatrix}
\mathbf{0}_{3\times 3}		&	\mathbf{Y}_{\tB}v	&	\mathbf{Y}_{\tW} v	&	\mathbf{G}_{S}v 	& 	\mathbf{G}_{T}v \\
\mathbf{Y}_{\tB}^{T}v	&	m_{\tB}				&	0				 	&	M_{\tB}			&	0	\\
\mathbf{Y}_{\tW}^{T}v	&	0					&	m_{\tW}				&	 0					&	M_{\tW}	\\
\mathbf{G}_{S}^{T}v	&	M_{\tB}			&		0				&	m_{S}				&	0	\\
\mathbf{G}_{T}^{T}v	&	0					&	M_{\tW}			& 	0				 	& 	m_{T}	
\end{pmatrix},
\label{eqn:numassmatrix}
\end{equation}
where
\begin{equation*}
\begin{aligned}
\mathbf{Y}_{\tB,\tW}^T &= \begin{pmatrix}
 Y_{\tB,\tW}^e &
 Y_{\tB,\tW}^\mu &
 Y_{\tB,\tW}^\tau
 \end{pmatrix}= \frac{M_{\tB,\tW}}{\Lambda_{M}}\begin{pmatrix}
 f^{e}_{\tB,\tW} &
 f^{\mu}_{\tB,\tW} &
 f^{\tau}_{\tB,\tW}
 \end{pmatrix}\, \equiv y_{\tB,\tW}\,\mathbf{u}_{\tB,\tW}^T\,,  \\
 \mathbf{G}_{S,T}^T &= \begin{pmatrix}
 G_{S,T}^e &
 G_{S,T}^\mu &
 G_{S,T}^\tau
 \end{pmatrix}~\, =\frac{\mGr}{\Lambda_{M}}\begin{pmatrix}
 d^{e}_{S,T} &
 d^{\mu}_{S,T} &
 d^{\tau}_{S,T}
 \end{pmatrix}\qquad\equiv g_{S,T}\,\mathbf{v}_{S,T}^T\,.
\end{aligned}
\end{equation*}
Here $\mathbf{u}_{\tB,\tW}$ and $\mathbf{v}_{S,T}$ are unit vectors, and the coefficients $y_{\tB,\tW}$ and $g_{S,T}$ read $M_{\tB,\tW}/\Lambda_M$ and $\mGr/\Lambda_M$, respectively. We ignore the \textit{CP}-violating phases that will emerge in \Cref{eqn:numassmatrix} as they will not affect our discussion. The elements of this mass matrix is highly hierarchical, $G_{S,T}\ll Y_{\tB,\tW}$ and $m_{\tB} \sim m_{\tW}\sim m_S\sim m_T \ll M_{\tB,\tW}$\,, solely due to the approximate $U(1)_{R-L}$ symmetry, determined by the hierarchy $\mGr \ll M_{\tB,\tW}$. Note that in the end, the Dirac gaugino masses will not be relevant for the neutrino masses.

In its most general form, the mass matrix in \Cref{eqn:numassmatrix} generates three massive light neutrinos with the correct mass splittings to match observations. Due to the numerous free parameters involved, it is not possible to analytically find the eigenvalues for the most general case. Therefore, we will focus on a simplified scenario as described below.

\textbf{Scenario 1:} We can first ignore the Majorana masses for the singlino and tripletino, setting $m_S = m_T =0$. In this case, the relevant Wilson coefficient for the dimension-5 Weinberg operator is
\begin{align}
    c^{d=5}&= \frac{y_{\tB} g_S}{\mbino} (\uc_{\tB} \vc_S^T+\vc_S \uc_{\tB}^T) + \frac{y_{\tW} g_T}{M_{\tW}} (\uc_{\tW} \vc_T^T+\vc_T \uc_{\tW}^T) \\
   &= \frac{m_{3/2}}{\Lambda_M^2}(\uc_{\tB} \vc_S^T+\vc_S \uc_{\tB}^T+\uc_{\tW} \vc_T^T+\vc_T \uc_{\tW}^T) \,.
\end{align}
One can set either one of $\mathbf{Y}_{\tB}, \mathbf{G}_S, \mathbf{Y}_{\tW}, \mathbf{G}_T $ to zero and still have two massive neutrinos with the correct mass splittings. Another simple case that results in one massless neutrino is when $\uc_{\tB}\propto \uc_{\tW}$ and/or $\vc_{\tB}\propto \vc_{\tW}$. (The light neutrino masses in these two cases are identical to \citep{Coloma:2016vod}.) This kind of linear dependence might point towards the UV completion of the terms introduced in \Cref{eqn:numassgenBW,eqn:RLviol}. We emphasize that, in its most general form, this scenario gives three massive neutrinos.

\textbf{Scenario 2:} We can ignore the $U(1)_{R-L}$-violating, dimension-5 terms given in \Cref{eqn:RLviol}, setting $\mathbf{G}_{S,T}=0$. Then the relevant Wilson coefficient becomes
\begin{align}
    c^{d=5}= -\frac{1}{\Lambda_{M}^2} \left (m_S\, \uc_{\tB} \uc_{\tB}^T + m_T\, \uc_{\tW} \uc_{\tW}^T \right )\,.
\end{align}
In this scenario, there is always one massless neutrino. Furthermore, the limit where $\uc_{\tB}\propto\uc_{\tW}$ results in two massless neutrinos and is not physical.

\begin{figure}[t]
\begin{center}
\includegraphics[width=.495 \linewidth]{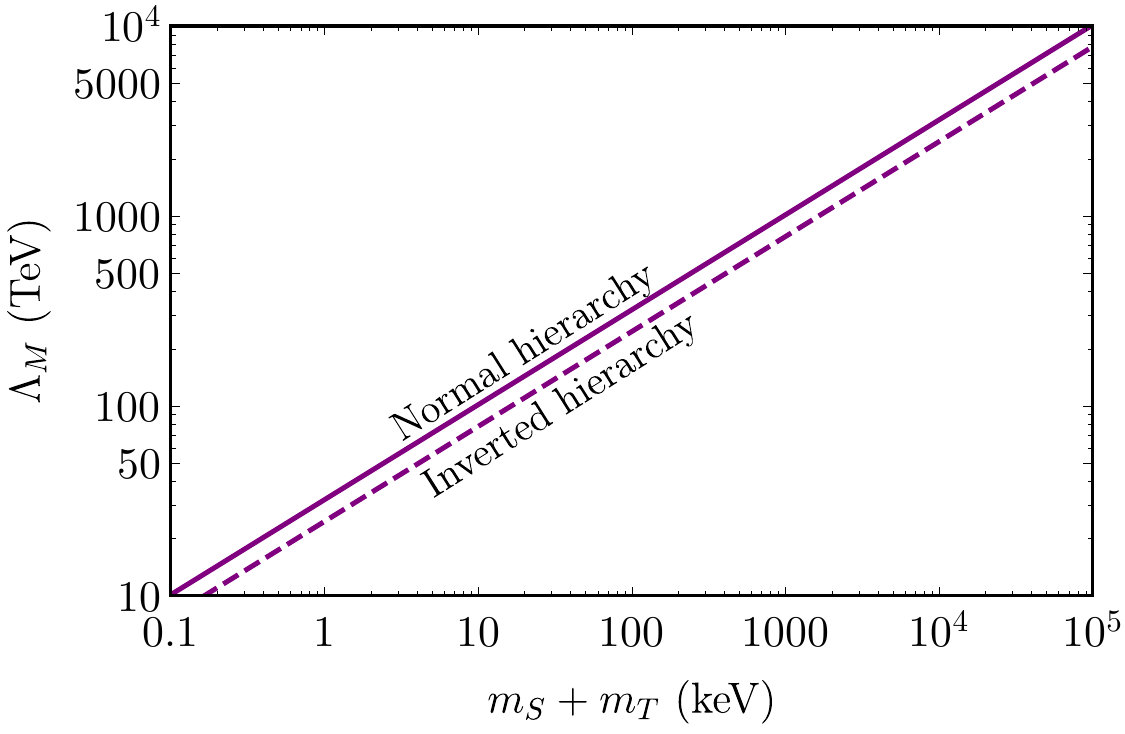}
\includegraphics[width=.495\linewidth]{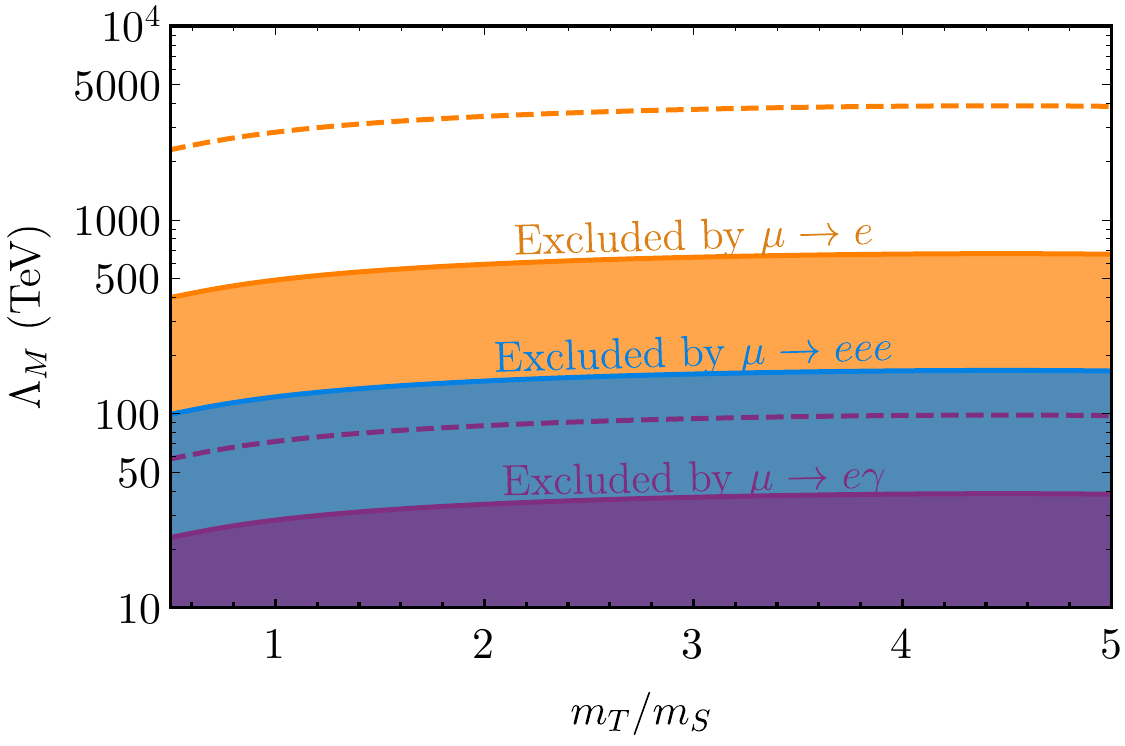}
\caption{\textbf{(Left)} Messenger scale $\Lambda_M$ vs. the sum of the singlet and triplet Majorana masses, $m_S+m_T$, that gives the correct neutrino masses for normal and inverted hierarchies. (The lightest neutrino is massless in this simplified scenario.) \textbf{(Right)} Constraints on the messenger scale $\Lambda_M$ for varying Majorana mass ratios $m_T/m_S$ for normal ordering. The purple shaded region is ruled out by the current $\mu \rightarrow e\gamma$ searches, ${\rm Br}(\mu\to e\gamma)<4.2\times 10^{-13}$ \citep{TheMEG:2016wtm}. The dashed purple line shows a projected limit for ${\rm Br}(\mu\to e\gamma)<10^{-14}$ \citep{Meucci:2022qbh}. The blue shaded region is excluded by ${\rm Br}(\mu\to eee) < 1.0\times 10^{-12}$~\citep{SINDRUM:1987nra}. The strongest constraints come from $\mu\to e$ conversion in gold, $R_{\mu\to e}<7\times 10^{-13}$~\citep{SINDRUM2006}, shown in orange. The orange dashed line shows the projected limit for $R_{\mu\to e}<6.2\times 10^{-16}$ expected from Mu2e \citep{Meucci:2022qbh, Mu2e:2022ggl} using aluminum. See text for details. For these plots, central values in \Cref{eqn:numixvariables} were used.}
\label{fig:numass}
\end{center}
\end{figure}

We analyze \textbf{Scenario 2} in further detail because it is analytically solvable and is fairly distinct from what has been studied in \citep{Coloma:2016vod}. (See \Cref{sec:appendix}.)  Setting $\mathbf{G}_{S,T}=0$, light neutrino masses in normal ordering (NO) are
\begin{equation}
\begin{aligned}        
m_1 = 0\,, \quad
m_{2,3}= \frac{v^2(m_S+m_T)}{\sqrt{2}\Lambda_{M}^2 } \sqrt{ 1 - 2\beta \mp \sqrt{1-4\beta}}\,,
\end{aligned}
\label{Eqn:numassNO}
\end{equation}
where $m_2<m_3$ and the parameter $\beta$ is set by the mass-squared splitting ratios \textit{r}, and have the following form for both mass ordering scenarios: 
\begin{equation}
\begin{aligned}
\beta_{\rm IO}&= \frac{\sqrt{r+1} \left(\sqrt{r+1}- 1\right)^2}{ r^2}\simeq 0.25\,, \quad \quad\quad\quad\quad\quad {\rm with} \quad r=\frac{|\Delta m^2_{\rm sol}|}{|\Delta m^2_{\rm atm}|}\simeq 0.03\,,\\
\beta_{\rm NO} &= -2r(r+1) + \sqrt{r (r+1)} (2 r+1)\simeq 0.13\,.
\end{aligned}
\label{Eqn:masssplit}
\end{equation}
We expect the singlet/triplet Majorana masses $m_{S,T}$ to be proportional to the gravitino mass $m_{3/2}$ up to $\octino(1)$ factors when $U(1)_{R-L}$ symmetry is approximately broken. We show in \Cref{Section:LFVconst} that low energy constraints require $\Lambda_M \gtrsim 500~$TeV, which translates to $m_S+m_T \sim \octino({\rm MeV})$. See \Cref{fig:numass}.

\begin{figure}[t]
\begin{center}
\includegraphics[width=.75 \linewidth]{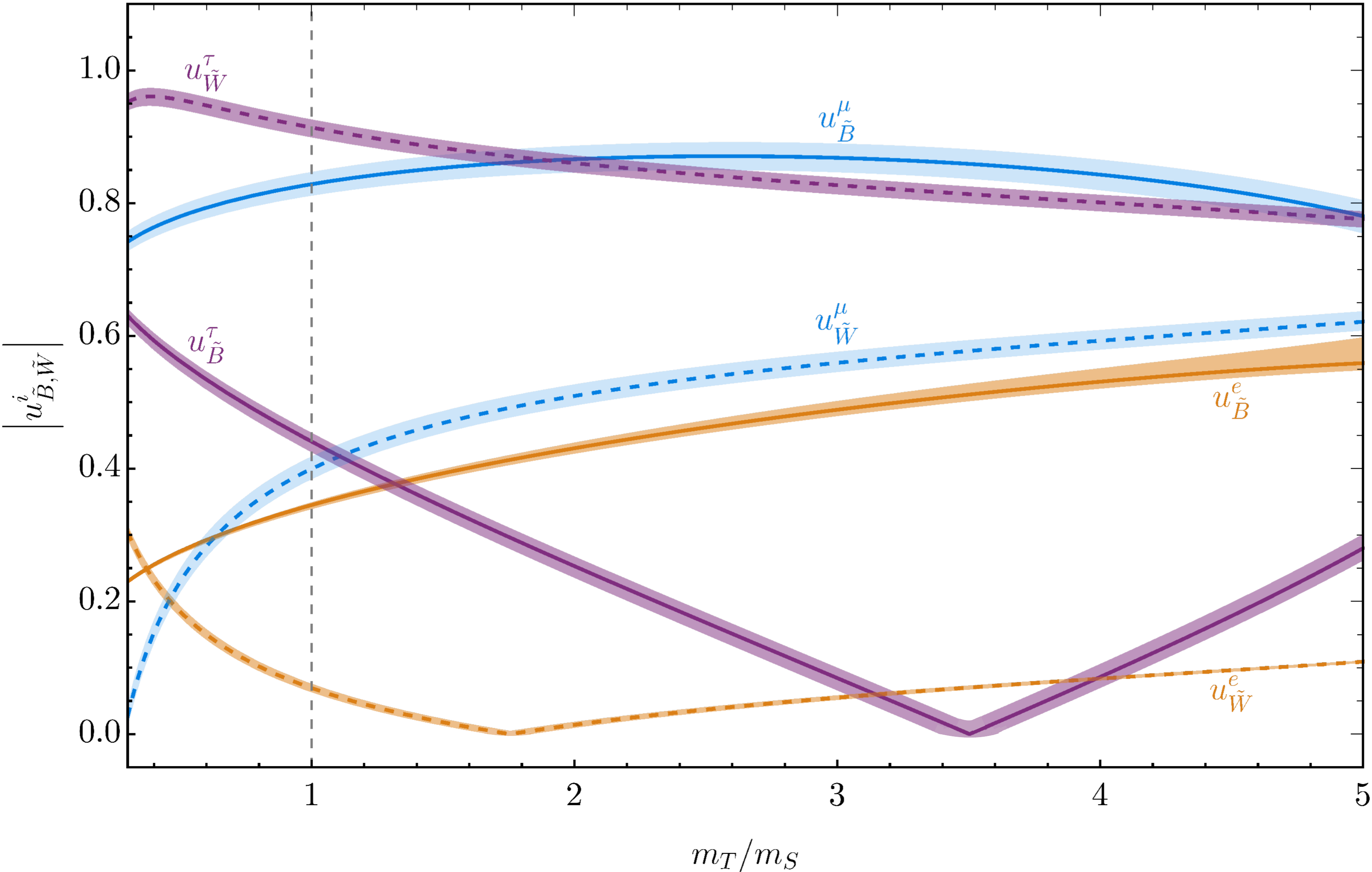}
\caption{The mixing matrix elements between light neutrinos and \bivo/\wivo, as defined in \Cref{eqn:numassmatrix}, in normal ordering. These are fixed by the PMNS matrix for a given value of $m_T/m_S$ where the bands correspond to varying the mixing angles within their $1\sigma$ measured values given in \Cref{eqn:numixvariables}. The case where $m_T=m_S$ matches the pure \bivo~scenario introduced in \citep{Coloma:2016vod}. There are accidental cancellations which drive $u_{\tW}^e$ and $u^\tau_{\tB}$ to zero for $m_T/m_S\sim 1.8$ and $\sim 3.5$ respectively.}
\label{fig:uBW}
\end{center}
\end{figure}

For a given Majorana mass ratio $m_T/m_S$, the PMNS matrix fixes the vectors $\mathbf{Y}_{\tB,\tW} = y_{\tB,\tW} \mathbf{u}_{\tB,\tW}$ up to the overall factors $y_{\tB,\tW}=M_{\tB,\tW}/\Lambda_M$. We show the mixing coefficients $u_{\tB,\tW}^i$, for $i=e,\mu,\tau$, in \Cref{fig:uBW} as a function of $m_T/m_S$ for normal ordering. There are accidental cancellations which drive $u_{\tW}^e$ and $u^\tau_{\tB}$ to zero for $m_T/m_S\sim 1.8$ and $\sim 3.5$ respectively. These sorts of cancellations can be interesting for future model building if, for example, one desires a sterile neutrino without significant mixing with the electron neutrino~\citep{Bertoni:2014mva}. This behavior will also affect the LHC signatures, which will be discussed in \Cref{Section:LHCpheno}. The gray dashed vertical line corresponds to $m_T=m_S$, which effectively reproduces the scenario introduced in \citep{Coloma:2016vod} with the identification $\mathbf{Y}_{\tW}\to \mathbf{G}$ in Equation (9) of \citep{Coloma:2016vod}. We note that there is an ambiguity between what is called \bivo~or \wivo~in these vectors. Namely, changing $m_S \leftrightarrow m_T$ and $u_{\tB}\leftrightarrow u_{\tW}$ leaves the neutrino phenomenology unchanged. See \Cref{sec:appendix} for details.

%%%%%%%%%%%%%%%%%%%%%%%%%%%%%%%%%%%
%%%%%% Low energy constraints %%%%%
%%%%%%%%%%%%%%%%%%%%%%%%%%%%%%%%%%%

\section{Low-energy Constraints}\label{Section:LFVconst}

\begin{figure}[t]
\begin{center}
\includegraphics[width=1.00 \linewidth]{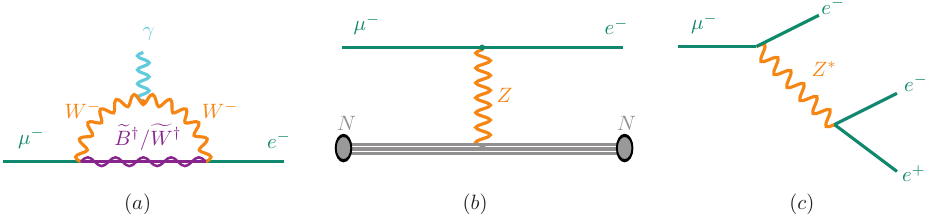}
\caption{Feynman diagrams for the lepton-flavor violating processes that dominantly constrain the messenger scale in our model. (\textit{a}) $\mu\to e\gamma$ at one-loop via $\tB/\tW$--neutrino mixing, (\textit{b}) $\mu-e$ conversion in nuclei and (\textit{c}) $\mu \to eee$ at tree level via chargino--charged lepton mixing.}
\label{fig:LFV-feynmandiagrams}
\end{center}
\end{figure}

The mixing of \bivo~and \wivo~with light neutrinos can result in observable lepton-flavor-violating (LFV) effects, which can be constrained by (non-)observations. This is particularly important in this work since \wivo~is an $SU(2)_L$-triplet in a type III seesaw model. As in generic type III seesaw models, $U(1)_{R-L}$-conserving terms in \Cref{eqn:numassgenBW} mixes charginos and charged leptons, inducing flavor-changing neutral and charged currents at tree level. Here, we summarize the most important constraints in this model. (See \Cref{fig:LFV-feynmandiagrams} for the processes we consider.)

Following the framework detailed in \citep{Abada:2007ux,Abada:2008ea}, one can constrain the (dimensionless) coefficient of the dimension-6 Weinberg operator,
\begin{align}
\epsilon^{d=6} = v^2 \left| \mathbf{Y}^T\frac{1}{\mathbf{\Lambda}^T \mathbf{\Lambda}} \mathbf{Y} \right|\,, \quad {\rm where} \quad \mathbf{Y} = \left(\mathbf{Y}_{\tB}\,, \mathbf{Y}_{\tW}\right), \quad \mathbf{\Lambda} = \begin{pmatrix}
    M_{\tB} & 0 \\
    0 & M_{\tW}
\end{pmatrix}\,, \label{eq:epsilon}
\end{align}
where the vectors $\mathbf{Y}_{\tB,\tW}$ are defined \Cref{eqn:numassmatrix} and their numerical values are given in \Cref{fig:uBW}. Note that while the dimension-5 operators that generate the light neutrino masses are suppressed by the small Majorana masses, the coefficient above is not necessarily suppressed, as expected in an ISS texture. Furthermore, since $\mathbf{Y}_{\tB,\tW} \propto M_{\tB,\tW}/\Lambda_M$, the coefficient $\epsilon^{d=6}\propto v^2/\Lambda_M^2$ is independent of the Dirac \bivo~and \wivo~masses. Consequently, the constraints will be only on the SUSY messenger scale $\Lambda_M$.

Detailed calculations of the constraints on type III seesaw models can be found in, e.g., \citep{Abada:2007ux,Abada:2008ea}. By far the strongest constraints are on the $e-\mu$ element of \Cref{eq:epsilon}:
\begin{align}
\left(\epsilon^{d=6}\right)_{e\mu} = \frac{v^2}{\Lambda_M^2}\left|u^e_{\tB}u^\mu_{\tB}+u^e_{\tW}u^\mu_{\tW}\right|\,.
\end{align}
As representative examples, we consider the following three lepton-flavor-violating processes. Final constraints are shown in the right panel of \Cref{fig:numass}.
\begin{enumerate}
    \item In this hybrid type I+III model, as in the pure type I case introduced in \citep{Coloma:2016vod}, $\mu\to e\gamma$ decay proceeds at one loop. The most recent limit on the branching fraction comes from MEG~\citep{TheMEG:2016wtm}, ${\rm Br}(\mu\to e \gamma)< 4.2\times 10^{-13}$. MEG II is expected to lower this limit to ${\rm Br}(\mu\to e \gamma)<  10^{-14}$. These translate into the constraints:
    \begin{align}
        \left(\epsilon^{d=6}\right)_{e\mu}^{\rm now} < 10^{-5}\,, \quad \quad \left(\epsilon^{d=6}\right)_{e\mu}^{\rm future} < 1.6 \times10^{-6}\,.
    \end{align}
    These limits are shown as the purple shaded region and the purple dashed line in the right panel of \Cref{fig:numass}. 
    \item Chargino mixing with the charged leptons generates an effective $Z\!-\!\mu\!-\!e$ vertex at tree level proportional to $\left(\epsilon^{d=6}\right)_{e\mu}$, which then allows for the tree level $\mu\to eee$ decay. The branching fraction of this process is constrained to be ${\rm Br}(\mu \to eee)<1.0\times 10^{-12}$~\citep{SINDRUM:1987nra}, which gives
     \begin{align}
        \left(\epsilon^{d=6}\right)_{e\mu} < 1.1\times 10^{-6}\,.
    \end{align}
    This constraint is shown as the blue shaded region in the right panel of \Cref{fig:numass}. It can be seen that this constraint will still be better than the forecasted $\mu\to e\gamma$ constraints in the near future. 
    \item The strongest constraint on the messenger scale comes from $\mu \to e$ conversion in nuclei. In contrast to the type I seesaw, here this process happens at tree level, facilitated by the same $Z\!-\!\mu\!-\!e$ vertex. The current best limit on this rate comes from the SINDRUM experiment using muonic gold: $R_{\mu e}<7\times 10^{-13}$~\citep{SINDRUM2006}. The proposed Mu2e experiment is expected to lower this limit by three orders of magnitude to $R_{\mu e} < 6.2\times 10^{-16}$~\citep{Mu2e:2022ggl}. The corresponding constraints 
     \begin{align}
        \left(\epsilon^{d=6}\right)_{e\mu}^{\rm now} < 3.4 \times10^{-8}\,, \quad \quad \left(\epsilon^{d=6}\right)_{e\mu}^{\rm future} < 1.0 \times10^{-9}\,,
    \end{align}
    are shown as the shaded orange region and the orange dashed line, respectively, in \Cref{fig:numass}.
\end{enumerate}

Overall, the constraints in this scenario require $\Lambda_M \gtrsim 500$~TeV, which is much stronger than the limits found in \citep{Coloma:2016vod}. This is due to the flavor-changing neutral currents (FCNC) induced by the chargino-charged lepton mixing at tree level. Note that every order of magnitude improvement in measurements raises the bound on $\Lambda_M$ by a factor of $10^{1/4}\sim 1.8$ since the rates for these processes are proportional to $1/\Lambda_M^4$.

%%%%%%%%%%%%%%%%%%%%%%%%%%%%%%%%%%%
%%%%%%%% LHC Pheno %%%%%%%%%%%%%%%%
%%%%%%%%%%%%%%%%%%%%%%%%%%%%%%%%%%%
\section{LHC Phenomenology} \label{Section:LHCpheno}
%%%%%%%%%%%%%%%%%%%%%%%%%%%%%%%%%%%

In most of the generic type I seesaw models, the SM-singlet RH neutrinos are not produced efficiently at the LHC since the only relevant interactions are generated through neutrino mixing. In contrast, the comparable minimal scenario introduced in \citep{Coloma:2016vod} has a rich collider phenomenology. In this model, The SM-singlet \emph{RH neutrino} is a bino, which can be produced in squark decays. Furthermore, the lightest neutralino, which is assumed to be pure bino, decays to SM particles via its mixing with the neutrinos. The lifetime depends on the bino mass and the SUSY messenger scale. These features generate unique signatures at ATLAS and CMS, as well as at proposed experiments like MATHUSLA, CODEX-b, and SHiP \citep{Gehrlein:2019gtk,Gehrlein:2021hsk,Ipek:2023jdp}. 

On the other hand, type III seesaw models employ EW triplets, which could have observable collider signatures via their EW interactions. The scenario we consider here automatically gives rise to a hybrid type I+III texture, where the SM-singlet \emph{RH neutrino} is the bino and the EW-triplet \emph{RH neutrino} is the wino. In this section, we comment on the LHC phenomenology of this scenario. 

\begin{figure}
     \centering
         \includegraphics[width=1.0 \linewidth]{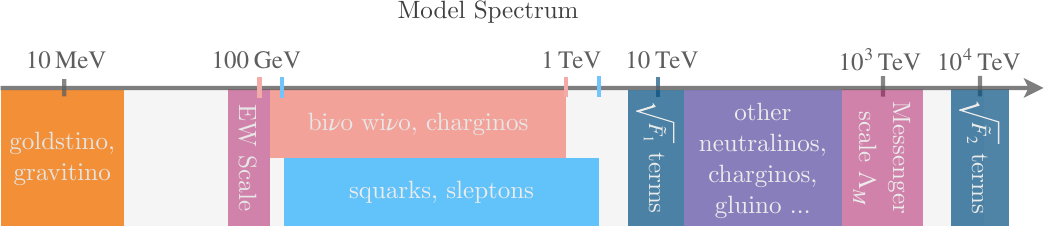}
     \caption{An example particle spectrum of the model described in \Cref{Section:model}.}
     \label{fig:spectrum}
 \end{figure}

As described in the previous section, low-energy LFV constraints require a messenger scale $\Lambda_M\sim \octino(1000~{\rm TeV})$ or larger. Consequently, to account for neutrino masses, the singlet/triplet Majorana masses require to be $\octino(10~{\rm MeV})$. Unlike the mechanism introduced in \citep{Coloma:2016vod}, neutrino masses do not directly translate into a requirement on the gravitino mass. However, we still expect the $U(1)_{R-L}$-breaking singlet/triplet Majorana masses to be proportional to the gravitino mass. We assume this proportionality holds up to an $\octino(1)$ constant. Therefore, the gravitino mass is expected to be around $\octino(10~{\rm MeV})$ as well. We assume that the gravitino is the lightest supersymmetric particle (LSP) and the two of the lightest neutralinos are purely bi$\nu$o- and \wivo-like. (In contrast to \citep{Gehrlein:2019gtk}, here wi$\nu$o is not decoupled.) Since we assume an approximately conserved $U(1)_R$ symmetry, the Majorana masses are much smaller than the Dirac masses for the gauginos. Hence we take the \bivo~and \wivo~masses to be $M_{\tB}, M_{\tW}$ respectively. We take charginos to be degenerate with the \wivo. We consider gluinos to be decoupled, but sfermions need not be decoupled. The mass spectrum we assume is shown in \Cref{fig:spectrum}.

%%%%%%%%%%%%%%%%%
\subsection{Bi$\nu$o as the lightest neutralino}
%%%%%%%%%%%%%%%%%%
%%%%%%%%%%%%%%%%%%
\subsubsection*{Bi$\nu$o phenomenology}
%%%%%%%%%%%%%%%%%%
When gluino is decoupled, but squarks are accessible at the LHC, \bivo~will be produced in squark decays in pair production of squarks with the cross section
\begin{align}
    \sigma_{\tB, \tB^\dagger} = {\rm Br}(\tilde{q}\to \tB q)^2\times \sigma(pp\to \tilde{q}\tilde{q}^\dagger)\,.
\end{align}
 If \bivo~is heavier than $\sim 100~$GeV, after being produced, it will decay to on-shell leptons, gauge bosons, and the Higgs through its mixing with the light neutrinos,
\begin{align}
    \tB \to W^-\ell^+\,,\quad \tB \to Z\bar{\nu}\,,\quad \tB \to h\bar{\nu}~, \label{eq:bivodecays}
\end{align}
as well to a gravitino and a photon, $\tB \to \tilde{G}\gamma$. This last process is suppressed by $M_{\rm Pl}^2$ and is not relevant for colliders. The \bivo ~lifetime $\Gamma_{\tB}\sim \sum\limits_i Y_i^2 M_{\tB}\sim M_{\tB}^3/\Lambda_M^2$ is $\octino(10~{\rm MeV})$ for a messenger scale $\Lambda_M\sim 100~$TeV. Hence, unlike generic MSSM scenarios, \bivo, which is the lightest neutralino, decays promptly at the LHC.\footnote{If the messenger scale is $\Lambda_M\gtrsim 10^8~$GeV or if \bivo ~is lighter than $\sim 80~$GeV, \bivo ~becomes long-lived on collider scales. The collider phenomenology of a long-lived \bivo~was studied in \citep{Gehrlein:2021hsk, Ipek:2023jdp}.} 

Bi$\nu$o signals at the LHC include final states with jets, leptons, and missing energy. In fact, ${\rm jets}+\slashed{E}_T$ final states have the highest branching fraction with $\sim 25\%$. In the previous LHC study of the scenario where neutrino masses are generated via solely \bivo~interactions \citep{Gehrlein:2019gtk}, the \wivo~was decoupled, leading to ${\rm Br}(\tilde{q}\to \tB q)=100\%$. Relevant LHC searches with $\sqrt{s}=13~$TeV and $\mathcal{L}=36~{\rm fb}^{-1}$ were recast to constrain the \bivo--squark  parameter space. It was shown that squarks as light as $350~$GeV are allowed for $M_{\tB}<150~$GeV and squarks above $950~$GeV were not constrained. 

When \wivo~mass is lighther than squarks, squark decays to \bivo s can effectively be ignored, and the analysis in \citep{Gehrlein:2019gtk} becomes irrelevant.The \bivo~production in \wivo~and chargino decays explained below will be important in this scenario.

%%%%%%%%%%%%%
\subsubsection*{Wi$\nu$o/chargino phenomenology}
%%%%%%%%%%%%%%
Since they are $SU(2)_L$-triplets, \wivo~and the charginos give rise to rich phenomenology at the LHC. Here we will highlight various scenarios to emphasize the novel signals of this type I+III inverse seesaw mechanism at the LHC.  

If the squarks are accessible at the LHC, \wivo ~and charginos are produced in their decays together with a quark. Charginos decay to $W^\pm \tB$ while \wivo ~decays to $Z\tB$, assuming kinematically allowed. In R-parity conserving MSSM, the lightest neutralino is stable and contributes to the signal as missing energy. Final states where gauge bosons decay hadronically and/or leptonically are considered depending on the search strategy, e.g. \citep{Aad:2021gluino,ATLAS:2021twp,ATLAS:2022zwa,CMS:2020bfa,CMS:2020cpy}.  When R-parity violation is allowed, \bivo ~can decay to a 3-body final state of leptons or quarks. We assume charginos are degenerate with the \wivo ~and that $M_{\tB} < M_{\tW} + M_Z$ such that \wivo~to \bivo~decays are kinematically allowed.

\begin{figure}[t]
\begin{center}
\includegraphics[width=.95 \linewidth]{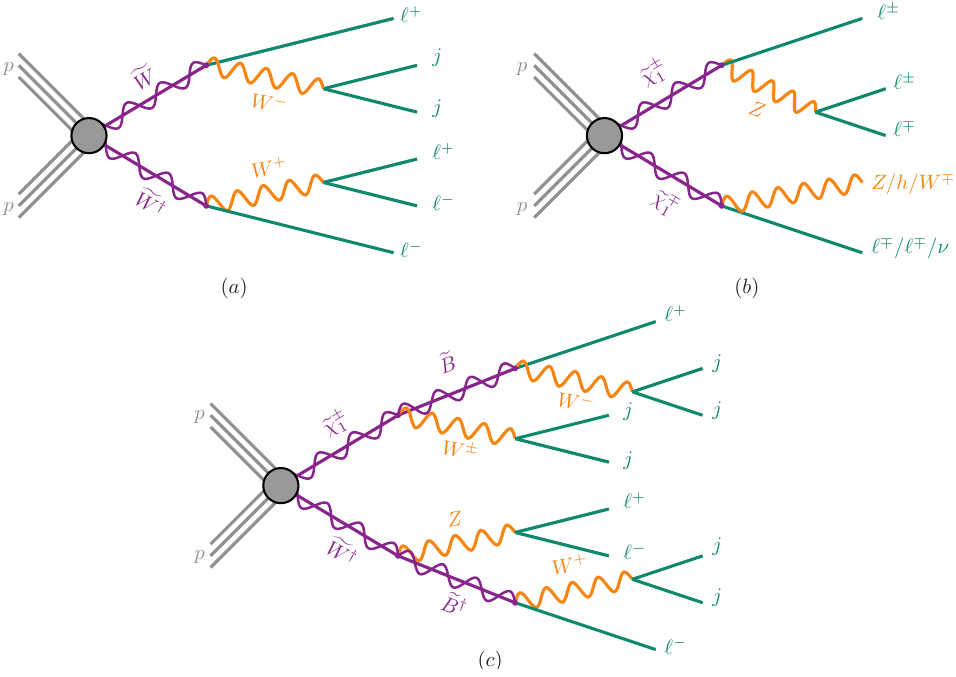}
\caption{Representitive LHC final states with leptons and jets. (\textit{a}) LSP \wivo ~pair production with decays mediated by $W^{\pm}$ boson. (\textit{b}) LSP chargino pair production with decays mediated by $Z$ boson and final states are associated with a weak gauge/Higgs boson and a lepton. (\textit{c}) \wivo ~and chargino pair production with decays mediated by LSP \bivo ~and weak gauge bosons.}
\label{fig:LHCdiagrams-combined}
\end{center}
\end{figure}

There are a few distinct features of the model we introduce here.
\begin{itemize}
    \item In addition to its normal decays to $Z\tB$, \wivo~can decay via mixing with the neutrinos; $\tW \to W^+\ell^-,~ Z\nu,~h\nu$. However, these channels are suppressed by the mixing angle $\theta^2\sim (y_{\tW}v/M_{\tW})^2\sim v^2/\Lambda_M^2\sim 10^{-6}$ compared to the $Z\tB$ channel. (See \Cref{fig:winodecay-feynmandiagrams}.)

     \begin{figure}[t]
     \centering
\includegraphics[width=0.3\linewidth]{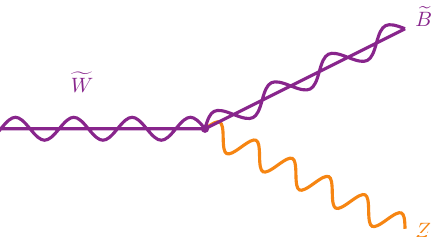}
\includegraphics[width=0.3\linewidth]{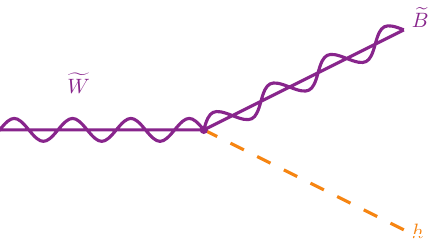}\\
\includegraphics[width=0.3\linewidth]{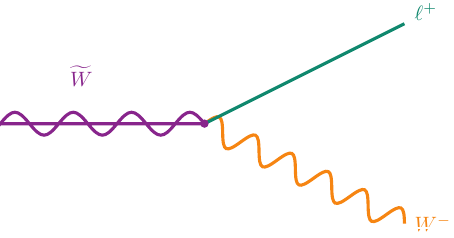}
\includegraphics[width=0.3\linewidth]{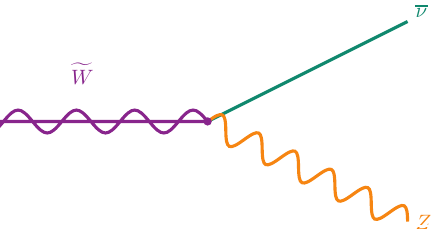}
\includegraphics[width=0.3\linewidth]{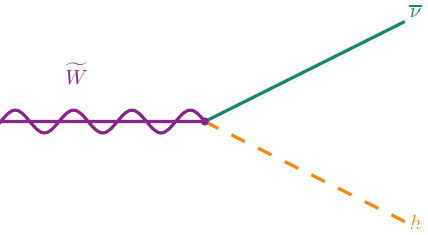}
\caption{If kinematically allowed, wi$\nu$o can decay to either a \bivo~or leptons and gauge bosons via mixing with the neutrinos. These latter decays are highly suppressed by the mixing angle.}
     \label{fig:winodecay-feynmandiagrams}
 \end{figure}
    
    \item Charginos can decay through the Lagrangian terms in \Cref{eqn:numassgenBW}, such as $\chi_1^\pm \to Z \ell^\pm$ induced by mixing with charged leptons. However, these mixing-induced decays are suppressed by $\theta^2\sim (y_{\tW}v/M_{\tW})^2\sim v^2/\Lambda_M^2\sim 10^{-6}$ and are irrelevant when \bivo~is the lightest neutralino, as $\chi_1^\pm \to W^\pm \tB$ decays dominate (see \Cref{fig:charginodecay-feynmandiagrams}). More important are flavor-violating $Z$ boson decays, e.g. $Z\to \mu e$ which is induced at tree level via this mixing.
     \begin{figure}[t]
     \centering
\includegraphics[width=0.3\linewidth]{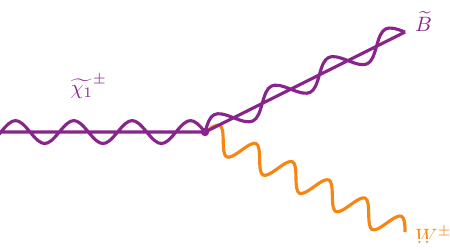}\\
\includegraphics[width=0.3\linewidth]{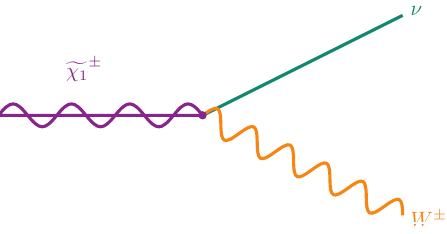}
\includegraphics[width=0.3\linewidth]{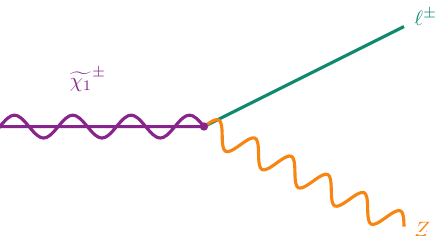}
\includegraphics[width=0.3\linewidth]{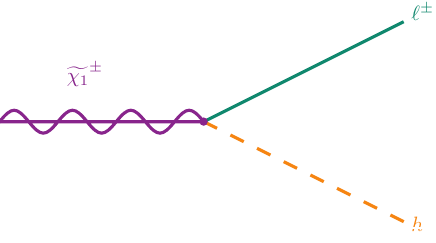}
\caption{For \bivo~LSP, the lightest chargino $\widetilde{\chi}_1^{\pm}$ will primarily decay to \bivo+$W^\pm$. However, charginos can also decay via their mixing with the charged leptons.}
     \label{fig:charginodecay-feynmandiagrams}
 \end{figure}
    
    \item Since \bivo~decays promptly via the decay channels given in \Cref{eq:bivodecays}, the signals will involve additional gauge bosons or the Higgs. (See \Cref{fig:LHCdiagrams-combined}.)
\end{itemize}

Ultimately, the signals can be similar to those expected from R-parity violating (RPV) decays, for example, through the $\lambda_{ijk}L_iL_jE_k^c$ term. (In RPV SUSY one can have $\chi_1^0\to \ell\ell \nu$ after sleptons are integrated out compared with $\tB\to Z(\to \ell \ell)\,\nu$ in this model.) Such a search has been carried out by ATLAS~\citep{ATLAS:2021yyr} with charged lepton final states. Wino neutralino is ruled out for $m(\chi_2^0)>1.6~$TeV for bino mass $m(\chi_1^0)>800~$GeV with an integrated luminosity of $139~{\rm fb}^{-1}$. We point out that although the final states are similar, the results of this and similar searches do not directly apply to our model. Firstly, due to the $U(1)_{R-L}$ symmetry, SUSY particle production cross-sections are generally small compared to MSSM. Furthermore, when \bivo~is heavier than the gauge bosons, the final state momentum distribution will be different than what is expected from 3-body decays in RPV SUSY. In fact, requiring the charged lepton pairs to reconstruct to $W/Z$ bosons could provide a smoking-gun signal for this scenario. Another important distinction is that the branching fractions to different lepton families ($e,\mu,\tau$) are determined by the observed neutrino mixing structure, as shown in \Cref{fig:uBW}. 

%%%%%%%%%%%%%%
\subsection{Wi$\nu$o as the lightest neutralino}
%%%%%%%%%%%%%%
If \wivo~is lighter than \bivo, \wivo~and charginos will only decay through the mixings induced by \Cref{eqn:RLviol} and explained in the previous section. (See also \Cref{fig:winodecay-feynmandiagrams,fig:charginodecay-feynmandiagrams}.) This scenario exactly matches the LHC phenomenology expected from a gauged $U(1)_{B-L}$ MSSM described in \citep{Dumitru:2018nct}. In this case, there is an ATLAS search for trilepton resonances~\citep{ATLAS:2020uer} with an integrated luminosity of $139~{\rm fb}^{-1}$ which directly applies to our model. (A decay channel would be $\chi_1^\pm \to \ell^\pm Z(\to \ell^\pm \ell^\mp)$ as shown in \Cref{fig:LHCdiagrams-combined}.) This search excludes wino/chargino masses between 100~GeV and 1.1~TeV, depending on their branching fraction to different lepton flavors, which is taken to be a free parameter for the analysis. In this search, the electron and muon final states are the most constraining. This observation is interesting when considering the \wivo-lepton mixing parameters shown in \Cref{fig:uBW}. For particular ratios of the singlet/triplet Majorana masses, $m_T/m_S$, it could be that chargino has no mixing with the electron while it mixes mostly with the tau, alleviating the constraints from this search. 

%%%%%%%%%%%%%%%%%%%%%%%%%%%%%%%%%
%%%%%%% Conclusions %%%%%%%%%%%%%
%%%%%%%%%%%%%%%%%%%%%%%%%%%%%%%%%

\section{Conclusion}
\label{sec:conclusions}

$U(1)_R$--symmetric MSSM conveniently alleviates the hierarchy problem without light stops. This is necessary given the stringent LHC constraints on stops and gluinos. When the global $U(1)_R$ symmetry is approximately broken by the gravitino mass, gauginos are expected to be pseudo-Dirac fermions with both Majorana and Dirac masses. We employed the pseudo-Dirac nature of the bino and the wino to generate light neutrino masses. To generate the neutrino masses, we extended the global symmetry to $U(1)_{R-L}$, and we introduced $U(1)_{R-L}$--conserving dim-5 operators and $U(1)_{R-L}$--violating dim-6 effective operators suppressed the messenger scale $\Lambda_M$.We found that this model can generate non-zero masses for all three neutrinos in its most general form. (This can be contrasted with the pure bino case in \citep{Coloma:2016vod}, where the lightest neutralino was predicted to be massless.) We then focused on a simplified scenario in which the only lepton-number violating terms are Majorana masses of the Dirac partners of the bino and the wino, $m_S$ and $m_T$ respectively. Then, the light neutrino masses are proportional to $(m_S+m_T)v^2/\Lambda_M^2$. The Majorana masses, which break $U(1)_{R-L}$, are expected to be proportional to the gravitino mass. The hierarchy between the gravitino mass and the messenger scale can explain the smallness of the neutrino masses. Interestingly, we found that for certain values of $m_S/m_T$, the bino or the wino does not mix with certain neutrino flavors. This property could be applied to other model-building scenarios where one needs a sterile neutrino to preferentially mix with, for example, tau neutrinos. Further work can be done by turning on all sources of lepton-number violation. In this most general case, a numerical scan of the large parameter space is needed.

The bino is a singlet, and the wino is a triplet under $SU(2)_L$. Consequently, the model described here generates a hybrid type I+III (inverse) seesaw mechanism.The  \wivo-neutrino mixing term also causes mixing between charginos and charged leptons, leading to tree-level FCNCs. We calculated the constraints coming from $\mu\to e \gamma$, $\mu\to eee$ and $\mu\to e$ conversion in nuclei and found that $\Lambda_M > (500-1000)$~TeV for $m_T/m_S \sim 0.2-5$. We also showed that future $\mu\to e$ conversion experiments like Mu2e \citep{Mu2e:2022ggl} can probe a messenger scale around 5000~TeV.

Finally, we discussed the rich LHC phenomenology expected from this model of neutrino mass generation. We assumed two mass orderings in the neutralino sector. \textbf{1)} If the lightest neutralino is a pure bino and the next-to-lightest neutralino is a pure wino, winos and charginos will decay to a bino and a gauge boson. In this case, the bino is short-lived and decays to leptons and a gauge boson or the Higgs. Hence, this model can generate final states with four electroweak gauge bosons. \textbf{2)} If the wino is the lightest neutralino, it can directly decay into leptons, gauge bosons, or Higgs via its mixing with the neutrinos. The branching fractions to specific leptons are predicted by the neutrino mixing parameters.

%%%%%%%%%%%%%%%%%%%%%%%%%%%%%%%%%
%%%%%%% Acknowledgements %%%%%%%%%%%%%
%%%%%%%%%%%%%%%%%%%%%%%%%%%%%%%%%

\section*{Acknowledgements}
This research was undertaken thanks in part to funding from the Canada First Research Excellence Fund through the Arthur B. McDonald Canadian Astroparticle Physics Research Institute, and support from the Natural Sciences and Engineering Research Council of Canada (NSERC) under grant number SAPIN-2022-00024.

%%%%%%%%%%%%%%%%%%%%%%%%%%%%%
%%%%%%%% Appendix %%%%%%%%%%%
%%%%%%%%%%%%%%%%%%%%%%%%%%%%%

\appendix
\section{Neutrino Mass Matrix Diagonalization}
\label{sec:appendix}
We follow \citep{Gavela:2009cd} for the neutrino mass matrix diagonalization. We parametrize the couplings as 
\begin{equation}
\mathbf{Y}_{\tB,\tW}^T  \equiv y_{\tB,\tW}\,\mathbf{u}_{\tB,\tW}^T\, ,\ \mathbf{G}_{S,T}^T  \equiv g_{S,T}\,\mathbf{v}_{S,T}^T\,.
\end{equation}
where $\mathbf{u}_{\tB,\tW}$ and $\mathbf{v}_{S,T}$ are unit vectors whose
inner products are parameterized as $ \mathbf{u}_{\tB}^{\dagger}\mathbf{u}_{\tW}=\mathbf{u}_{\tW}^{\dagger}\mathbf{u}_{\tB}\equiv \lambda $. Coefficients $y_{\tB,\tW}$ and $g_{S,T}$ are given in the text in \Cref{sec:neutrinomasses}. We ignore the \textit{CP}-violating phases in \Cref{eqn:numassmatrix}.

As we discuss in \Cref{sec:neutrinomasses}, we focus on a simplified scenario (\textbf{Scenario 2}) in which the couplings $G_{S,T}$ are taken to be zero. In this case, the Wilson coefficient for the dimension-5 Weinberg operator reads 
\begin{align}
    c^{d=5}= -\frac{\mGr}{\Lambda_{M}^2} \left (\kappa_S\, \uc_{\tB} \uc_{\tB}^T + \kappa_T\, \uc_{\tW} \uc_{\tW}^T \right )\equiv -\frac{\mGr}{\Lambda_{M}^2} \octino ,
    \label{eqn:s2-dim5Wein}
\end{align}
where the coefficients $\kappa_{S,T}$ are defined in \Cref{eqn:STMajoranamass}. The resulting eigensystem is analytically solvable. Note that \Cref{eqn:s2-dim5Wein} is invariant under $m_S \leftrightarrow m_T$ and $\uc_{\tB}\leftrightarrow \uc_{\tW}$. This peculiar symmetry between \bivo ~and \wivo ~couplings does not change the neutrino phenomenology. Since $\octino$ is a symmetric complex matrix, the PMNS matrix $U_{\rm PMNS}$, diagonalizes the hermitian matrix $ \octino^\dagger \octino$: 
\begin{equation}
    \frac{\mGr^2 v^4}{\Lambda_M^4}U^\dagger_{\rm PMNS} \octino^\dagger \octino U_{\rm PMNS}=\octino_d^2,
    \label{eqn:diagonalization}
\end{equation}
where $\octino_d^2$ is the diagonal mass-squared matrix whose elements are the eigenvalues of $ \octino^\dagger \octino$. Therefore, the columns of the PMNS matrix can be written in terms of the eigenvectors of \Cref{eqn:s2-dim5Wein}: 
\[
U_{\rm PMNS}= \begin{pmatrix}
    U_{i 1}&U_{i 2}&U_{i 3}
\end{pmatrix}=\begin{pmatrix}
\hat{\mathbf{e}}_0&\hat{\mathbf{e}}_-&\hat{\mathbf{e}}_+    
\end{pmatrix}\, ,\quad i=e,\mu,\tau.
\]
 We focus on the normal mass-ordering scenario. Note that it is straightforward to generalize the eigensystem for both mass orderings. The eigenvalues in the normal ordering are found to be 
\begin{equation}
\begin{aligned}        
m_0 = 0, \,
m_{\pm}= \frac{v^2\mGr}{\sqrt{2}\Lambda_{M}^2 } \left [
\kappa_S^2+\kappa_T^2+2\kappa_S\kappa_T\lambda_{\rm NO}^2 \pm (\kappa_S+\kappa_T)\sqrt{ (\kappa_S-\kappa_T)^2 + 4\kappa_S \kappa_T \lambda_{\rm NO}^2}\right]^{1/2}.
\end{aligned}
\end{equation}
Here, $\lambda_{\rm NO}$ is parametrized as
\[ 
    \lambda_{\rm NO}=\sqrt{1+\beta_{\rm NO}\frac{(\kappa_S+\kappa_T)^2}{\kappa_S \kappa_T}},
\]
and it is set by the mass-squared splitting ratios, \Cref{Eqn:masssplit}. We assume the massive eigenstates have the form, 
\begin{equation}
    \hat{\mathbf{e}}_{\pm}=N_\pm(a_{\pm}\uc_{\tB}+b_{\pm}\uc_{\tW}),
\end{equation}
where the coefficients $a_{\pm}$, $b_{\pm}$ and $N_{\pm}$ can be determined by solving the eigenvalue problem, $\octino^\dagger \octino \hat{\mathbf{e}}_{\pm}=m_{\pm}^2\hat{\mathbf{e}}_{\pm}$:
\begin{align*}
        & a_{\pm}=-2\kappa_S \lambda_{\rm NO}\, ,\\ 
        & b_{\pm}=(\kappa_S-\kappa_T) \mp\sqrt{(\kappa_S-\kappa_T)^2+ 4\kappa_S \kappa_T \lambda_{\rm NO}^2}\, ,\\
        & N_{\pm}=\frac{1}{\sqrt{ a_{\pm}^2+b_{\pm}^2+2a_{\pm}b_{\pm}\lambda_{\rm NO}}}. 
\end{align*}
The entries in the PMNS matrix fix the mass eigenstates to accommodate the correct mixing structure. Hence, the vectors $u_{\tB,\tW}^{i}$ in the normal ordering are found to be
\begin{equation}
\begin{aligned}
       u_{\tB}^{i}&=\left(\frac{a_{-}}{b_{-}}-\frac{a_{+}}{b_{+}}\right)^{-1}\left[\frac{1}{b_{-}N_{-}}U_{i2}-\frac{1}{b_{+}N_{+}}U_{i3} \right], \\
        u_{\tW}^{i}&=\left(\frac{b_{-}}{a_{-}}-\frac{b_{+}}{a_{+}}\right)^{-1}\left[\frac{1}{a_{-}N_{-}}U_{i2}-\frac{1}{a_{+}N_{+}}U_{i3} \right]. 
\end{aligned}
\label{eqn:eigenvectors}
\end{equation}
The behavior of the vectors $u_{\tB,\tW}^{i}$ is given in \Cref{fig:uBW} as a function of $m_S/m_T=\kappa_S/\kappa_T$. In particular, for $\kappa_S=\kappa_T\equiv \kappa$, we recover the scenario in \citep{Coloma:2016vod}:
\begin{equation}
\begin{aligned}
         a_{\pm}&= \left.\pm b_{\pm} \right |_{\kappa_S=\kappa_T}=-2\kappa \lambda_{\rm NO}\quad ,\\ 
         N_{\pm}&= \left.\frac{1}{\sqrt{2}|a_{\pm}|\sqrt{ 1\mp \lambda_{\rm NO}}}\right |_{\kappa_S=\kappa_T}, \\
         u_{\tB}^{i} &=\frac{1}{\sqrt{2}}\left[\sqrt{1+\lambda_{NO}}U_{i3}+\sqrt{1-\lambda_{NO}} U_{i2} \right] , \\
         v_{S}^i &=\frac{1}{\sqrt{2}}\left[\sqrt{1+\lambda_{NO}}U_{i3}-\sqrt{1-\lambda_{NO}} U_{i2} \right]\, . 
\end{aligned}
\label{eqn:eigensysrecover}
\end{equation}
Using the mixing parameters in \Cref{eqn:numixvariables} in \Cref{eqn:eigensysrecover}, we retrieve the exact values in Equation (9) in \citep{Coloma:2016vod}:
\begin{equation}
    \uc_{\tB}=\begin{pmatrix}
        0.35\\ 0.85 \\ 0.39
    \end{pmatrix}\ {\rm and\ }     \vc_{S}=\begin{pmatrix}
        -0.06\\ 0.44 \\ 0.89
    \end{pmatrix}.
\end{equation}

%\singlespace
\bibliographystyle{JHEP}
\bibliography{nuref}{}
\end{document}